\documentclass[10pt,conference]{IEEEtran}

\IEEEoverridecommandlockouts

\usepackage{balance}

\usepackage{cite}
\usepackage{amsmath,amssymb,amsfonts}
\usepackage{algorithmic}
\usepackage{graphicx}
\usepackage{subfigure}
\usepackage{textcomp}
\usepackage{xcolor}

\usepackage{booktabs}
\usepackage{caption}

\usepackage{multirow} 
\usepackage{array}
\usepackage{makecell}
\usepackage{url}
\usepackage[shortlabels]{enumitem}

\usepackage[colorlinks,
            linkcolor=red,
            anchorcolor=blue,
            citecolor=green]{hyperref}

\newcommand{\starchat}{StarChat-$\beta$}
\newcommand{\codellama}{CodeLlama-Instruct}

\usepackage{pifont}
\usepackage{tcolorbox}
\newcommand{\finding}[1]{
\begin{center}
\begin{tcolorbox}[colback=white!15, colframe=black, boxsep=-0.15cm, middle=-0.15cm]
\textbf{\ding{43} Finding}
$\blacktriangleright$
{#1}
$\blacktriangleleft$
\end{tcolorbox}
\end{center}
}
\newcommand{\summary}[1]{
\begin{center}
\begin{tcolorbox}[colback=gray!15, colframe=black, boxsep=-0.15cm, middle=-0.15cm]
\textbf{\ding{46} Summary}
$\blacktriangleright$
{#1}
$\blacktriangleleft$
\end{tcolorbox}
\end{center}
}

\usepackage{etoolbox}
\makeatletter
\patchcmd{\authornote}{\g@addto@macro\addresses{\@authornotemark}}{}{}{}
\makeatother

\usepackage[normalem]{ulem}

\def\BibTeX{{\rm B\kern-.05em{\sc i\kern-.025em b}\kern-.08em
    T\kern-.1667em\lower.7ex\hbox{E}\kern-.125emX}}
    
\begin{document}

\title{Source Code Summarization in the Era of Large Language Models}

\author{
\IEEEauthorblockN{Weisong Sun$^{1}$, Yun Miao$^2$, Yuekang Li$^3$, Hongyu Zhang$^4$, Chunrong Fang$^{2}$, Yi Liu$^1$, Gelei Deng$^1$, \\ Yang Liu$^1$, Zhenyu Chen$^2$} 
\IEEEauthorblockA{
    $^1$College of Computing and Data Science, Nanyang Technological University Singapore, Singapore\\
    $^2$State Key Laboratory for Novel Software Technology, Nanjing University, Nanjing, China\\
    $^3$School of Computer Science and Engineering, University of New South Wales, Sidney, Australia \\
    $^4$School of Big Data and Software Engineering, Chongqing University, Chongqing, China\\
    weisong.sun@ntu.edu.sg, miaoyun001my@gmail.com, yuekang.li@unsw.edu.au, hyzhang@cqu.edu.cn, \\fangchunrong@nju.edu.cn,
    yi009@e.ntu.edu.sg, 
    gelei.deng@ntu.edu.sg yangliu@ntu.edu.sg, zychen@nju.edu.cn
    }
}

\maketitle

\thispagestyle{plain} 

\begin{abstract}
To support software developers in understanding and maintaining programs, various automatic (source) code summarization techniques have been proposed to generate a concise natural language summary (i.e., comment) for a given code snippet. Recently, the emergence of large language models (LLMs) has led to a great boost in the performance of code-related tasks. 
In this paper, we undertake a systematic and comprehensive study on code summarization in the era of LLMs, which covers multiple aspects involved in the workflow of LLM-based code summarization. 
Specifically, we begin by examining prevalent automated evaluation methods for assessing the quality of summaries generated by LLMs and find that the results of the GPT-4 evaluation method are most closely aligned with human evaluation. 
Then, we explore the effectiveness of five prompting techniques (zero-shot, few-shot, chain-of-thought, critique, and expert) in adapting LLMs to code summarization tasks. Contrary to expectations, advanced prompting techniques may not outperform simple zero-shot prompting. 
Next, we investigate the impact of LLMs' model settings (including top\_p and temperature parameters) on the quality of generated summaries. We find the impact of the two parameters on summary quality varies by the base LLM and programming language, but their impacts are similar. 
Moreover, we canvass LLMs' abilities to summarize code snippets in distinct types of programming languages. The results reveal that LLMs perform suboptimally when summarizing code written in logic programming languages compared to other language types (e.g., procedural and object-oriented programming languages). 
Finally, we unexpectedly find that \codellama{} with 7B parameters can outperform advanced GPT-4 in generating summaries describing code design rationale and asserting code properties. 
We hope that our findings can provide a comprehensive understanding of code summarization in the era of LLMs. 
\end{abstract}

\begin{IEEEkeywords}
large language model, source code summarization, prompt engineering
\end{IEEEkeywords}

\section{Introduction}
\label{sec:introduction}
Code comments are vital for enhancing program comprehension~\cite{1981-Comments-on-Program-Comprehension} and facilitating software maintenance~\cite{2005-Documentation-Essential-Software-Maintenance}. 
While it is considered good programming practice to write high-quality comments, the process is often labor-intensive and time-consuming~\cite{2005-Documentation-Essential-Software-Maintenance, 2020-CPC, 2023-EACS}. 
As a result, high-quality comments are frequently absent, mismatched, or outdated during software evolution, posing a common problem in the software industry~\cite{2015-How-do-Developers-Document, 2019-Code-Comment-Inconsistencies, 2022-Practitioners-Expectations-on-Comment-Generation}. 
Automatic code summarization (or simply, code summarization), a hot research topic~\cite{2022-Evaluation-Neural-Code-Summarization, 2023-SIDE, 2023-Prompt-CS}, addresses this challenge by developing advanced techniques/models for automatically generating natural language summaries (i.e., comments) for code snippets, such as Java methods or Python functions, provided by developers.

\begin{figure}[!t]
  \centering
  \includegraphics[width=1.0\linewidth]{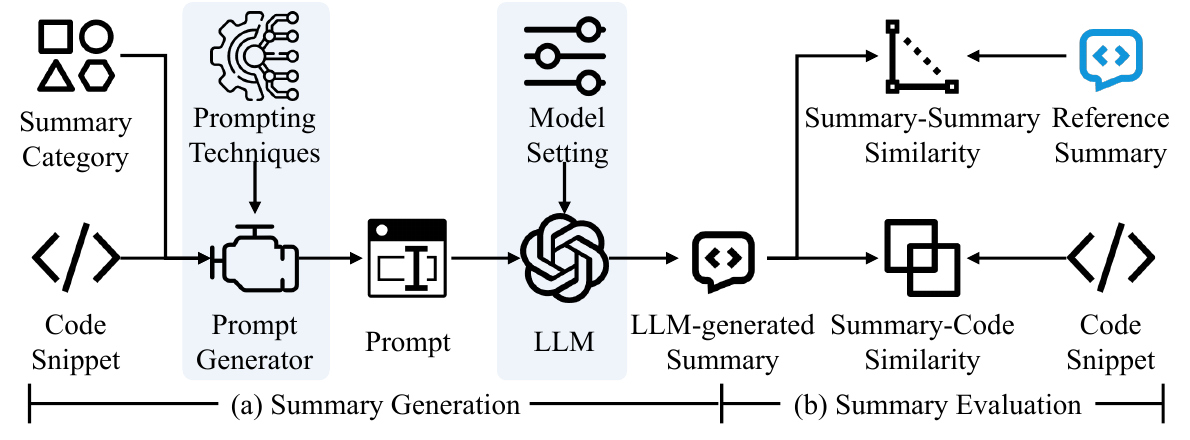}
  \caption{General workflow of LLM-based code summarization and its effectiveness evaluation}
  \vspace{-6mm}
  \label{fig:LLM-based_code_summarization}
\end{figure}

Recently, with the success of large language models (LLMs) in natural language processing (NLP)~\cite{2023-Shortcut-Learning-of-LLM-in-NLP}, an increasing number of software engineering (SE) researchers have started integrating them into the resolution process of various SE tasks~\cite{2023-LLMs-for-SE-Survey, 2023-LLMs-for-SE-Review}. In this study, we focus on the application of LLMs on the code summarization tasks. 
Figure~\ref{fig:LLM-based_code_summarization} shows the general workflow of LLM-based code summarization and its effectiveness evaluation methods. 
In the summary generation process, the input consists of a code snippet and the expected summary category. The input is passed to a prompt generator equipped with various prompt engineering techniques (referred to as prompting technique), which constructs a prompt based on input. This prompt is then used to instruct LLMs to generate a summary of the expected type for the input code snippet.
In the summary evaluation process, a common method used to automatically assess the quality of LLM-generated summaries is to compute the text or semantic similarity between the LLM-generated summaries and the reference summaries.

There have been several recent studies investigating the effectiveness of LLMs in code summarization tasks~\cite{2022-Few-shot-Training-LLMs-for-Code-Summarization, 2022-No-More-Fine-tuning-in-Code-Intelligence, 2023-Automatic-Code-Summarization-via-ChatGPT, 2024-LLM-Few-Shot-Summarizers-Multi-Intent-Comment-Generation, 2024-Effectively-Tuning-Pre-trained-Code-Models}. 
These studies can help subsequent researchers rapidly understand the aspects of code summarization garnering attention in the era of LLMs, but they still have some limitations.
First, most of them only focus on one prompting technique, while some advanced prompting techniques have not been investigated and compared (e.g., chain-of-thought prompting). 
For example, Sun et al.~\cite{2023-Automatic-Code-Summarization-via-ChatGPT} solely focus on zero-shot prompting, while several other studies~\cite{2022-Few-shot-Training-LLMs-for-Code-Summarization, 2023-What-Makes-Good-In-Context-Demonstrations, 2024-LLM-Few-Shot-Summarizers-Multi-Intent-Comment-Generation} only focus on few-shot prompting. 
Second, they overlook the impact of the model settings (i.e., parameter configuration) of LLMs on their code summarization capabilities. There is no empirical evidence showing LLMs remain well in all model settings. 
Last but not least, these studies follow prior code summarization studies~\cite{2023-EACS, 2021-SiT, 2018-TL-CodeSum} to evaluate the quality of summaries generated by LLMs through computing text similarity (e.g., BLEU~\cite{2002-BLEU}, METEOR~\cite{2005-METEOR}, and ROUGE-L~\cite{2004-ROUGE}) or semantic similarity (e.g., SentenceBERT-based cosine similarity~\cite{2022-Semantic-Metrics-for-Evaluating-Code-Summarization}) between the LLM-generated summaries and the reference summaries, detailed in Section~\ref{subsec:answer_to_RQ1}. However, prior research by Sun et al.~\cite{2023-Automatic-Code-Summarization-via-ChatGPT} has shown that compared to traditional code summarization models, the summaries generated by LLMs significantly differ from reference summaries in expression and tend to describe more details. Consequently, whether these traditional evaluation methods are suitable for assessing the quality of LLM-generated summaries remains unknown. 

To address these issues, in this paper, we conduct a systematic study on code summarization in the era of LLMs, which covers various aspects involved in the LLM-based code summarization workflow. 
Considering that the choice of evaluation methods directly impacts the accuracy and reliability of the evaluation results, we first systematically investigate the suitability of existing automated evaluation methods for assessing the quality of summaries generated by LLMs (including \codellama{}, \starchat{}, GPT-3.5, and GPT-4).  
Specifically, we compare multiple automated evaluation methods (including methods based on summary-summary text similarity, summary-summary semantic similarity, and summary-code semantic similarity) with human evaluation to reveal their correlation. Inspired by the work in NLP~\cite{2023-ChatGPT-NLG-Evaluator, 2023-Disinformation-Capabilities-of-LLMs, 2023-G-Eval}, we also explore the possibility of using the LLMs themselves as evaluation methods. The experimental results show that \textbf{among all automated evaluation methods, the GPT-4-based evaluation method overall has the strongest correlation with human evaluation}. 
Second, we conduct comprehensive experiments on three widely used programming languages (Java, Python, and C) datasets to explore the effectiveness of five prompting techniques (including zero-shot, few-shot, chain-of-thought, critique, and expert) in adapting LLMs to code summarization tasks. The experimental results show that the optimal choice of prompting techniques varies for different LLMs and programming languages. \textbf{Surprisingly, the more advanced prompting techniques expected to perform better may not necessarily outperform simple zero-shot prompting.} For instance, when the base LLM is GPT-3.5, zero-shot prompting outperforms the other four more advanced prompting techniques overall on three datasets.   
Then, we investigate the impact of two key model settings/parameters, including top\_p and temperature, on LLMs' code summarization performance. These two parameters may affect the randomness of generated summaries. The results demonstrate that \textbf{the effect of top\_p and temperature on summary quality varies depending on the base LLM and programming language}. As alternative parameters, they exhibit a similar impact on the quality of LLM-generated summaries. 
Furthermore, unlike existing studies that simply experimented with multiple programming languages, we reveal the differences in the code summarization capabilities of LLMs across five types (including procedural, object-oriented, scripting, functional, and logic programming languages) encompassing ten programming languages: Java, Python, C, Ruby, PHP, JavaScript, Go, Erlang, Haskell, and Prolog. The Erlang, Haskell, and Prolog datasets are built by ourselves and we make them public to the community. We find that \textbf{across all five types of programming languages, LLMs consistently perform the worst in summarizing code written in logic programming languages}. 
Finally, we investigate the ability of LLMs to generate summaries of different categories, including \texttt{What}, \texttt{Why}, \texttt{How-to-use-it}, \texttt{How-it-is-done}, \texttt{Property}, and \texttt{Others}. 
The results reveal that the four LLMs perform well in generating distinct categories of summaries. For example, \textbf{\codellama{} excels in generating \texttt{Why} and \texttt{Property} summaries, while GPT-4 is good at generating \texttt{What}, \texttt{How-it-is-done}, and \texttt{How-to-use} summaries}. 
Our comprehensive research findings will assist subsequent researchers in quickly and deeply understanding the various aspects involved in the workflow of code summarization based on LLMs, as well as in designing advanced LLM-based code summarization techniques for specific fields.

In summary, we make the following contributions.
\begin{itemize}
    \item To the best of our knowledge, we conduct the first investigation into the feasibility of applying LLMs as evaluators to assess the quality of LLM-generated summaries.

    \item We conduct a thorough study of code summarization in the era of LLMs, covering multiple aspects of the LLM-based code summarization workflow, and come up with several novel and unexpected findings and insights. These findings and insights can benefit future research and practical usage of LLM-based code summarization.

    \item We make our dataset and source code publicly accessible~\cite{2024-LLM4CodeSummarization} to facilitate the replication of our study and its application in extensive contexts.
\end{itemize}

\section{Background and Related Work}
\label{sec:background_and_related_work}
Code summarization is the task of automatically generating natural language summaries (also called comments) for code snippets. Such summaries serve various purposes, including but not limited to explaining the functionality of code snippets~\cite{2020-Hybrid-DeepCom, 2022-Practitioners-Expectations-on-Comment-Generation, 2024-LLM-Few-Shot-Summarizers-Multi-Intent-Comment-Generation}. The research on code summarization can be traced back to as early as 2010 when Sonia Haiduc et al.~\cite{2010-Program-Comprehension-with-Code-Summarization} introduced automated text summarization technology to summarize source code. Later on, following the significant success of neural machine translation (NMT) research in the field of NLP~\cite{2014-NMT, 2014-On-the-Properties-of-NMT}, a large number of researchers migrate its underlying encoder-decoder architecture to code summarization tasks~\cite{2020-Transformer-based-Approach-for-Code-Summarization, 2020-Code-to-Comment-Translation, 2020-Rencos, 2022-Evaluation-Neural-Code-Summarization, 2023-EACS}.  
In the past two years, research on LLM-based code summarization has mushroomed. 
Fried et al.~\cite{2023-InCoder} introduce an LLM called InCoder, and try zero-shot training on the CodeXGLUE~\cite{2021-CodeXGLUE} Python dataset. InCoder achieves impressive results, but fine-tuned small PLMs like CodeT5 can still outperform the zero-shot setting. 
Ahmed et al.~\cite{2022-Few-shot-Training-LLMs-for-Code-Summarization} investigate the effectiveness of few-shot prompting in adapting LLMs to code summarization and find that it can make Codex significantly outperform fine-tuned small PLMs (e.g., CodeT5). 
Given the concern of potential code asset leakage when using commercial LLMs (e.g., GPT-3.5), Su et al.~\cite{2024-Distilled-GPT-for-Code-Summarization} utilize knowledge distillation technology to distill small models from LLMs (e.g., GPT-3.5). Their experimental findings reveal that the distilled small models can achieve comparable code summarization performance to LLMs. 
Gao et al.~\cite{2023-What-Makes-Good-In-Context-Demonstrations} investigate the optimal settings for few-shot learning, including few-shot example selection methods, few-shot example order, and the number of few-shot examples.  
Geng et al.~\cite{2024-LLM-Few-Shot-Summarizers-Multi-Intent-Comment-Generation} investigate LLMs' ability to address multi-intent comment generation. 
Ahmed et al.~\cite{2024-Semantic-Augmentation-of-Prompts-for-Code-Summarization} propose to enhance few-shot samples with semantic facts automatically extracted from the source code. 
Sun et al.~\cite{2023-Automatic-Code-Summarization-via-ChatGPT} design several heuristic questions to collect the feedback of ChatGPT, thereby finding an appropriate prompt to guide ChatGPT to generate in-distribution code summaries. 
Rukmono et al.~\cite{2023-Achieving-High-Level-Software-Component-Summarization} address the unreliability of LLMs in performing reasoning by applying a chain-of-thought prompting strategy. 
Recently, some studies~\cite{2022-No-More-Fine-tuning-in-Code-Intelligence, 2023-CodePrompt, 2023-Prompt-CS} have also investigated the applicability of Parameter-Efficient Fine-Tuning (PEFT) techniques in code summarization tasks. 
In this paper, we focus on uncovering the effectiveness of various prompting techniques in adapting LLMs to code summarization without fine-tuning.

\section{Study Design}
\label{sec:study_desigh}

\subsection{Research Questions}
\label{subsec:research_questions}

\noindent This study aims to answer the following research questions:

\textbf{RQ1: What evaluation methods are suitable for assessing the quality of summaries generated by LLMs?} 
Existing research on LLM-based code summarization~\cite{2022-Few-shot-Training-LLMs-for-Code-Summarization, 2023-What-Makes-Good-In-Context-Demonstrations, 2024-LLM-Few-Shot-Summarizers-Multi-Intent-Comment-Generation} widely follow earlier studies~\cite{2018-TL-CodeSum, 2022-Semantic-Metrics-for-Evaluating-Code-Summarization} and employ automated evaluation metrics (e.g., BLEU) to evaluate the quality of LLM-generated summaries. 
However, recent studies~\cite{2023-Automatic-Code-Summarization-via-ChatGPT, 2024-Distilled-GPT-for-Code-Summarization} have shown that LLM-generated summaries surpass reference summaries in quality. Therefore, evaluating LLM-generated summaries based on their text or semantic similarity to reference summaries may not be appropriate. 
This RQ aims to discover a suitable method for automated assessment of the quality of LLM-generated summaries.

\textbf{RQ2: How effective are different prompting techniques in adapting LLMs to the code summarization task?} 
This RQ aims to unveil the effectiveness of several popular prompting techniques (e.g., few-shot and chain-of-thought) in adapting LLMs to code summarization tasks. 

\textbf{RQ3: How do different model settings affect LLMs' code summarization performance?} 
To better meet diverse user needs, LLMs typically offer configurable parameters (i.e., model settings) that allow users to control the randomness of model behaviour. In this RQ, we adjust the randomness of the generated summaries by modifying LLMs' parameters and see the impact of different model settings on the performance of LLMs in generating code summaries.

\textbf{RQ4: How do LLMs perform in summarizing code snippets written in different types of programming languages?}
Programming languages are diverse in types (e.g., object-oriented and functional programming languages), with their implementations of the same functional requirements being similar or entirely different. The scale of programs implemented with them in Internet/open-source repositories also varies, which may result in differences in the mastery of knowledge of these languages by LLMs. Hence, this RQ aims to reveal the differences in LLMs' capabilities to summarize code snippets across diverse programming language types.

\textbf{RQ5: How do LLMs perform on different categories of summaries?}
Previous research~\cite{2020-CPC, 2021-Why-My-Code-Summarization-Not-Work, 2023-DOME} has shown that summaries can be classified into various categories according to developers' intentions, including \texttt{What}, \texttt{Why}, \texttt{How-to-use-it}, \texttt{How-it-is-done}, \texttt{Property}, and others.
Therefore, in this RQ, we aim to explore the ability of LLMs to generate summaries of different categories.

\subsection{Experimental LLMs}
\label{subsec:experimental_LLMs}
\noindent We select four LLMs as experimental representatives.

\textbf{\codellama{}.} Code Llama~\cite{2023-CodeLlama} is a family of LLMs for code based on Llama 2~\cite{2023-llama2}. 
It provides multiple flavors to cover a wide range of applications: foundation models, Python specializations (Code Llama-Python), and instruction-following models (Code Llama-Instruct) with 7B, 13B, 34B, and 70B parameters. In this study, we evaluate the summaries generated by \codellama{} 7B. We also verify the ability of \codellama{} 70B to act as an evaluator in RQ1.

\textbf{\starchat{}.} \starchat{}~\cite{2023-starchat} is an LLM with 16B parameters fine-tuned on StarCoderPlus~\cite{2023-starcoderplus}. 
Compared with StarCoderPlus, \starchat{} excels in chat-based coding assistance.

\textbf{GPT-3.5.} GPT-3.5~\cite{2022-OpenAI} is an LLM provided by OpenAI. It is trained with massive texts and codes. 
It can understand and generate natural language or code. 

\textbf{GPT-4.} GPT-4 is an improved version of GPT-3.5, which can solve difficult problems with greater accuracy. OpenAI has not disclosed the specific parameter scale of GPT- 3.5 and GPT-4. Our study uses gpt-3.5-turbo and gpt-4-1106-preview.

\textit{Model Settings.} Apart from RQ3 where we investigate the impact of model settings, we uniformly set the temperature to 0.1 to minimize the randomness of LLM's responses and highlight the impact of evaluation methods/prompting techniques/programming language types/summary categories. Note that the LLM-generated summaries are relatively long, and usually consist of several sentences. By observing some examples, we find that the first sentences are suitable as the final summaries, and the following sentences elaborate on some details and supplementary explanations. Therefore, we extract the first sentences as the final summaries.

\subsection{Prompting Techniques}
\label{subsec:prompting_techniques}
\noindent We compare five commonly used prompting techniques below.

\textbf{Zero-Shot.} Zero-shot prompting adapts LLMs to downstream tasks using simple instructions. In our scenario, the input to LLMs consists of a simple instruction and a code snippet to be summarized. We expect LLMs to output a natural language summary of the code snippet. Therefore, we follow~\cite{2023-Automatic-Code-Summarization-via-ChatGPT} and adopt the input format: \emph{Please generate a short comment in one sentence for the following function: $\langle$code$\rangle$}. 

\textbf{Few-Shot.} Few-shot prompting (also known as in-context learning~\cite{2024-LLM-Few-Shot-Summarizers-Multi-Intent-Comment-Generation, 2023-What-Makes-Good-In-Context-Demonstrations}) provides not only straightforward instruction but also some examples when adapting LLMs to downstream tasks. The examples serve as conditioning for subsequent examples where we would like LLMs to generate a response. In our scenario, the examples are pairs of \emph{$\langle$code snippet, summary$\rangle$}. According to the findings of Gao et al.~\cite{2023-What-Makes-Good-In-Context-Demonstrations}, we set the number of examples to 4 to achieve a balance between LLMs' performance and the cost of calling the OpenAI API.

\textbf{Chain-of-Thought.} Chain-of-thought prompting adapts LLMs to downstream tasks by providing intermediate reasoning steps~\cite{2022-Chain-of-Thought-Prompting}. These steps enable LLMs to possess complex reasoning capabilities. In this study, we follow Wang et al.~\cite{2023-Element-aware-Summarization-with-LLMs} and apply chain-of-thought prompting to the code summarization task through the following four steps: 
\begin{enumerate}[(1)]
    \item Instruction 1: Input the code snippet and five questions about the code in the format\\ \textit{``Code: \textbackslash n$\langle$code$\rangle$\\ Question: \textbackslash n$\langle$Q1$\rangle$\textbackslash n$\langle$Q2$\rangle$\textbackslash n$\langle$Q3$\rangle$\textbackslash n$\langle$Q4$\rangle$\textbackslash n$\langle$Q5$\rangle$\textbackslash n''}
    
    \item Get LLMs' response to Instruction 1, i.e., Response 1.
    
    \item Instruction 2: \textit{``Let's integrate the above information and generate a short comment in one sentence for the function.''}
    
    \item Get LLMs' response to Instruction 2, i.e., Response 2. Response 2 contains the comment generated by LLMs for the code snippet.
\end{enumerate}
when asking Instruction 2, Instruction 1 and Response 1 are paired as history prompts and answers and input into the LLM. 

\textbf{Critique.} 
Critique prompting improves the quality of LLMs' answers by asking LLMs to find errors in the answers and correct them. We follow Kim et al.~\cite{2023-Critique} and perform critique prompting on the code summarization task through the six steps below: 
\begin{enumerate}[(1)]
    \item Instruction 1: Similar to zero-shot prompting, input the instruction and the code snippet in the format \emph{``Please generate a short comment in one sentence for the following function: \textbackslash n$\langle$code$\rangle$''}
    
    \item Get LLMs' response to Instruction 1, i.e., Response 1. Response 1 contains the temporary comment generated by LLMs for the code snippet.
    
    \item Instruction 2: \emph{``Review your previous answer and find problems with your answer.''}
    
    \item Get LLMs' response to Instruction 2, i.e., Response 2.
    
    \item Instruction 3: \emph{``Based on the problems you found, improve your answer.''}
    
    \item Get LLMs' response to Instruction 3, i.e., Response 3. Response 3 contains the modified comment, which is the final comment of the code snippet.
\end{enumerate}
when prompting each instruction, previous instructions and responses are fed into the LLMs as pairs of history prompts and answers. 

\textbf{Expert.} Expert prompting first asks LLMs to generate a description of an expert who can complete the instruction (e.g., through few-shot prompting), and then the description serves as the system prompt for zero-shot prompting. 
We use the few-shot examples provided by Xu et al.~\cite{2023-Expert-Prompting} and employ few-shot prompting to let LLMs generate a description of an expert who can ``Generate a short comment in one sentence for a function.'' This description will replace the default system prompt of LLMs. By default, we use the system prompt~\cite{2023-System-Prompt-from-codellama} of \codellama{} for all LLMs to ensure fairness in comparison. Then, we utilize the same steps as zero-shot prompting to adapt LLMs to generate summaries.

\subsection{Experimental Datasets}
\label{subsec:experimental_datasets}

\noindent The sources of the datasets utilized in our experiments include:

\textbf{CodeSearchNet (CSN).} The CodeSearchNet corpus~\cite{2019-CodeSearchNet-Challenge} is a vast collection of methods accompanied by their respective comments, written in Go, Java, JavaScript, PHP, Python, and Ruby. 
This corpus has been widely used in studying code summarization~\cite{2022-UniXcoder, 2023-EACS, 2023-Adapter-Tuning-Code-Search-and-Summarization}. 
We use the clean version of the CSN corpus provided by Lu et al.~\cite{2021-CodeXGLUE} in CodeXGLUE. 
We randomly select 200 samples for each programming language from the test set of this corpus for experiments.

\textbf{CCSD.} The CCSD dataset is provided by Liu et al.~\cite{2020-CCSD}. They crawl data from 300+ projects such as Linux and Redis. The dataset contains 95,281 $\langle$function, summary$\rangle$ pairs. Similarly, we randomly select 200 samples from the final dataset for experiments.

In addition to the above two sources, we construct three new language datasets to evaluate LLM's code summarization capabilities across more programming language types.

\textbf{Erlang, Haskell, and Prolog Datasets.} Erlang and Haskell are Functional Programming Languages (FP), and Prolog belongs to Logic Programming Languages (LP). 
To construct the three datasets, we sort the GitHub repositories whose main language is Erlang/Haskell/Prolog according to the number of stars, and crawl data from the top 50 repositories. Following Husian et al.~\cite{2019-CodeSearchNet-Challenge}, 
(1) we remove any projects that do not have a license or whose license does not explicitly permit the re-distribution of parts of the project. 
(2) We consider the first sentence in the comment as the function summary. 
(3) We remove data where functions are shorter than three lines or comments containing less than 3 tokens. 
(4) We remove functions whose names contain the substring ``test''. 
(5) We remove duplicates by comparing the Jaccard similarities of the functions following Allamanis et al.~\cite{2019-The-Adverse-Effects-of-Code-Duplication}. Finally, we get 7,025/6,759/1,547 pairs of $\langle$function, summary$\rangle$ pairs. For each language, we randomly select 200 samples for experiments.

\begin{table}[!t]
    \caption{Datasets. PP: Procedural Programming Languages, OOP: Object-Oriented Programming Languages, SP: Scripting Programming Languages, FP: Functional Programming Languages, LP: Logic Programming Languages. }
    \label{tab:dataset_information}
    \centering  
    \scriptsize
    \tabcolsep=13pt
    \begin{tabular}{cccc}
        \toprule
        Language & Source & Type & Usage \\
        
        \midrule
        
        Java & CSN & OOP & RQ1, RQ2, RQ3, RQ4, RQ5 \\

        Python & CSN & SP & RQ1, RQ2, RQ3, RQ4 \\

        C & CCSD & PP & RQ1, RQ2, RQ3, RQ4 \\

        Ruby & CSN & SP & RQ4 \\
        
        PHP & CSN & SP & RQ4 \\
        
        Go & CSN & PP & RQ4 \\

        JavaScript & CSN & SP & RQ4 \\

        Erlang & by us & FP & RQ4 \\

        Haskell & by us & FP & RQ4 \\

        Prolog & by us & LP & RQ4 \\
        
        \bottomrule
        
    \end{tabular}
    \vspace{-4mm}
\end{table}

All in all, our experiments involve 10 programming languages across 5 types. Note that considering that experiments with LLMs are resource-intensive (especially those involving GPT, which are quite costly), not all experiments are conducted on all 10 programming language datasets. 
Specifically, we first conduct experiments associated with RQ1 and RQ2 on commonly used programming languages, including Java, Python, and C. Analyzing the results of these two RQs helps find a suitable automated evaluation method and a suitable prompting technique. Subsequent experiments for other RQs can be built upon these findings, thereby significantly reducing experimental costs. 
We use all 10 programming languages in the experiments for RQ4. 
In the experiments for RQ5, we only use the Java dataset because other programming languages lack readily available comment classifiers. While training such classifiers would be valuable, it falls outside the scope of this paper and is left for future exploration.

\section{Results and Findings}

\subsection{RQ1: What evaluation methods are suitable for assessing the quality of summaries generated by LLMs?}
\label{subsec:answer_to_RQ1}

\noindent1) \textit{Experimental Setup.}

\noindent\textbf{Comparison Evaluation Methods.} Existing automated evaluation methods for code summarization can be divided into the following three categories.

\textit{i. Methods based on summary-summary text similarity} assess the quality of the generated summary by calculating the text similarity between the generated summary and the reference summary. This category of methods is the most widely used in existing code summarization research~\cite{2023-EACS, 2022-Few-shot-Training-LLMs-for-Code-Summarization, 2023-What-Makes-Good-In-Context-Demonstrations, 2024-LLM-Few-Shot-Summarizers-Multi-Intent-Comment-Generation}. The text similarity metrics involved include BLEU, METEOR, and ROUGE-L, which compare the count of n-grams in the generated summary against the reference summary. 
The scores of BLEU, METEOR, and ROUGE-L are in the range of [0, 1]. 
The higher the score, the closer the generated summary approximates the reference summary, indicating superior code summarization performance. All scores are computed by the same implementation provided by~\cite{2020-Rencos}.

\textit{ii. Methods based on summary-summary semantic similarity} evaluate the quality of the generated summary by computing the semantic similarity between the generated summary and the reference summary. Existing research~\cite{2022-Semantic-Metrics-for-Evaluating-Code-Summarization} demonstrates that semantic similarity-based methods can effectively alleviate the issues of word overlap-based metrics, where not all words in a sentence have the same importance and many words have synonyms. In this study, we compare four such methods, including BERTScore~\cite{2020-BERTScore}, SentenceBert with Cosine Similarity (SBCS), SentenceBert with Euclidean Distance (SBED), and Universal Sentence Encoder~\cite{2018-USE} with Cosine Similarity (USECS). 
They are commonly used in code summarization studies~\cite{2021-Reassessing-Metrics-for-Code-Summarization, 2022-Semantic-Metrics-for-Evaluating-Code-Summarization, 2022-Retcom}. 
BERTScore~\cite{2020-BERTScore} uses a variant of BERT~\cite{2019-BERT} (we use the default $RoBERTa_{large}$) to embed every token in the summaries, and computes the pairwise inner product between tokens in the reference summary and generated summary. Then it matches every token in the reference summary and the generated summary to compute the precision, recall, and $F_1$ measure. In our experiment, we report the $F_1$ measure of BERTScore. The other three methods use a pre-trained sentence encoder (SentenceBert~\cite{2019-SentenceBERT} or Universal Sentence Encoder~\cite{2018-USE}) to produce vector representations of two summary sentences, and then compute the cosine similarity or euclidean distance of the vector representations. 
SBCS, SBED, and USECS range within [-1,1]. Higher values of SBCS and USECS represent greater similarity, while lower values of SBED indicate greater similarity.

\textit{iii. Methods based on summary-code semantic similarity} assess the quality of the generated summary by computing the semantic similarity between the generated summary and the code snippet to be summarized. 
Unlike the first two methods, this type of evaluation method does not rely on reference summaries and can effectively avoid issues related to low-quality and outdated reference summaries. SIDE proposed by Mastropaolo et al.~\cite{2023-SIDE} is a representative of this type of method. It is based on contrastive learning and has been trained to assess the relevance of a given textual summary for a Java method. Note that it has not been trained on other language datasets. Hence, in our experiment, SIDE is only used to evaluate the Java dataset. 
SIDE provides a continuous score ranging within [-1,1], where a higher value represents greater similarity. 
We present the scores reported by the above similarity-based evaluation methods in percentage.

\textit{Human Evaluation.} 
We conduct human evaluations as a reference for automated evaluation methods. Comparing the correlation between the results of automated evaluation methods and human evaluation can facilitate achieving the goal of this RQ, which is to find a suitable automated method for assessing the quality of LLM-generated summaries.  
To do so, we invite 15 volunteers (including 1 PhD candidate, 5 masters, and 9 undergraduates) with more than 3 years of software development experience and excellent English ability to carry out the evaluation. 
For each sample, we provide volunteers with the code snippet, the reference summary, and summaries generated by four LLMs, where the reference summary and the summaries generated by four LLMs are mixed and out of order. In other words, for each sample, volunteers do not know whether it is a reference or a summary generated by a certain LLM. We follow Shi et al.~\cite{2022-Evaluation-Neural-Code-Summarization} and ask volunteers to rate the summaries from 1 to 5 based on their quality where a higher score represents a higher quality. The final score of the summaries is the average of scores rated by 15 volunteers.

\begin{figure}[t]
  \centering
  \includegraphics[width=\linewidth]{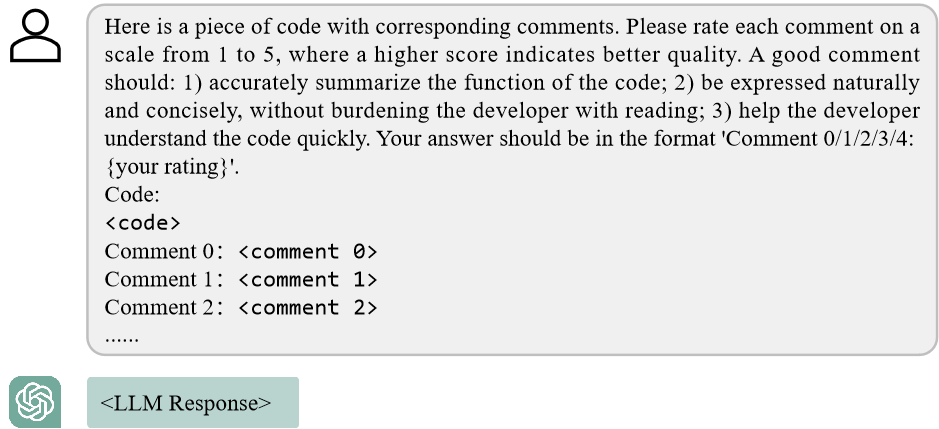}
  \caption{An example of using an LLM as an evaluator.}
  \label{fig:llm-eval}
  \vspace{-2mm}
\end{figure}

\textit{LLM-based evaluation methods.} 
Inspired by recent work in NLP~\cite{2023-ChatGPT-NLG-Evaluator, 2023-Disinformation-Capabilities-of-LLMs, 2023-G-Eval}, we also investigate the feasibility of employing LLMs as evaluators. Its advantage is that it does not rely on the quality of reference summaries, and the evaluation steps can be the same as human evaluation. 
Specifically, similar to human evaluation, when using LLMs as evaluators, for each sample, we input the code snippet to be summarized, the reference summary, and LLM-generated summaries, and ask LLMs to rate each summary from 1 to 5 where a higher score represents a higher quality of the summary. The specific prompt when using LLMs as evaluators is shown in Figure~\ref{fig:llm-eval}.

\noindent\textbf{Datasets and Prompting Techniques.}
In this RQ, to reduce the workload of human evaluation volunteers, we randomly select 50 samples from the Java, Python, and C datasets, respectively, which means 150 samples in total. We employ few-shot prompting to adapt the four LLMs to generate summaries for code snippets as recent studies~\cite{2022-Few-shot-Training-LLMs-for-Code-Summarization, 2024-LLM-Few-Shot-Summarizers-Multi-Intent-Comment-Generation,2023-What-Makes-Good-In-Context-Demonstrations} have demonstrated the effectiveness of this prompting technique on code summarization tasks.

\noindent2) \textit{Experimental Results.}

\noindent\textbf{Human Evaluation Results.}
Table~\ref{table:RQ1-human-evaluation-results} shows the human evaluation scores for reference summaries and summaries generated by the four LLMs. Observe that the scores of reference summaries in the three datasets are between 3 and 3.5 points, suggesting that the quality of the reference summaries is not very high. 
Therefore, evaluation methods based on summary-summary similarity may not accurately assess the quality of LLM-generated summaries.

Among the four LLMs, GPT-4 has the highest scores on the Java and C datasets, and GPT-3.5 attains the highest score on the Python dataset. This suggests that the quality of summaries generated by GPT-3.5 and GPT-4 is relatively high. 

\finding{According to human evaluation, the quality of reference summaries in the existing datasets is not particularly high. Summaries from general-purpose LLMs (e.g., GPT-3.5) excel over those from specialized code LLMs (e.g., \codellama{}) in quality.}

\begin{table}[t]
    \caption{Human evaluation scores for reference and LLM-generated summaries. The value in parentheses represents the percentage increase or decrease relative to the score of the corresponding reference summary.}
    \label{table:RQ1-human-evaluation-results}
    \centering
    \scriptsize
    \tabcolsep=7pt
    \begin{tabular}{lccc}
        \toprule
        \multirow{2}{*}{Summary from} & \multicolumn{3}{c}{Human Evaluation Score} \\

        \cmidrule{2-4}

        & Java & Python & C \\

        \midrule

        Reference & 3.19 & 3.56 & 3.05 \\

        \midrule

        \codellama & 3.93 (+23.20\%) & 3.88 (+8.99\%) & 4.15 (+36.07\%) \\

        \starchat{} & 3.18 (-0.31\%) & 3.14 (-11.80\%) & 3.49 (+14.43\%) \\

        GPT-3.5 & 4.00 (+25.39\%) & \textbf{4.16 (+16.85\%)} & 4.06 (+33.11\%) \\ 

        GPT-4 & \textbf{4.17 (+30.72\%)} & 4.06 (+14.04\%) & \textbf{4.25 (+39.34\%)} \\
        
        \bottomrule
        
    \end{tabular}
\end{table}

\noindent\textbf{Automated Evaluation Results.} 
Table~\ref{table:RQ1-all-metrics} displays the scores of the LLM-generated summaries reported by three categories of automated evaluation methods, and LLM-based evaluation methods. Observed that among the three methods based on summary-summary text similarity, 1) the BLEU-based and ROUGE-L-based methods give \starchat{} the highest scores on all three datasets; 2) the METEOR-based method gives \starchat{} the highest score (i.e., 18.19) on the Java dataset, while gives \codellama{} the highest scores (i.e., 21.64 and 17.29) on the Python and C datasets. 
Among the four methods based on summary-summary semantic similarity, BERTScore, SBCS, and SBED give the best scores to \starchat{}, and USECS gives the best score of 50.69 to \codellama{} on the Java dataset. On the Python and C datasets, the four methods consistently give the best scores to \codellama{} and \starchat{}, respectively. 
The summary-code semantic similarity-based method SIDE gives the highest score (i.e., 80.46) to \starchat{} on the Java dataset. 
On the Java dataset, the five LLM-based methods consistently give the highest scores to GPT-4. While on the Python dataset, they consistently award the highest scores to GPT-3.5. On the C dataset, \starchat{} gives the highest score to \codellama{} while other LLMs give the highest score to GPT-4. 

\finding{According to automated evaluation, overall, methods based on summary-summary text/semantic similarity tend to give higher scores to specialized code LLMs \starchat{} and \codellama{}, while LLM-based evaluators tend to give higher scores to general-purpose LLMs GPT-3.5 and GPT-4. The summary-code semantic similarity-based method tends to give higher scores to \starchat{} on the Java dataset.}

\begin{table*}[t]
    \caption{Automated evaluation scores for reference and LLM-generated summaries. S-S Tex.Sim.: methods based on summary-summary text similarity; S-S Sem.Sim.: methods based on summary-summary semantic similarity; S-C Sem.Sim.: methods based on summary-code semantic similarity. CodeLlama-I: \codellama{}. We bold the best score in each column.}
    \label{table:RQ1-all-metrics}
    \centering
    \scriptsize
    \tabcolsep=3pt
    \resizebox{1.0\linewidth}{!}{
    \begin{tabular}{cccccccccccccccc}
        \toprule

        \multirow{2}{*}{Language} & \multirow{2}{*}{\makecell{Summary\\from}} & \multicolumn{3}{c}{S-S Tex.Sim.} & \multicolumn{4}{c}{S-S Sem.Sim.} & \multicolumn{1}{c}{S-C Sem.Sim.} & \multicolumn{5}{c}{LLM-based Evaluation Method} & \multirow{2}{*}{Human} \\

        \cmidrule(lr){3-5}\cmidrule(lr){6-9}\cmidrule(lr){10-10}\cmidrule(lr){11-15}

        & & BLEU & METEOR & ROUGE-L & BERTScore & SBCS & SBED & USECS & SIDE & CodeLlama-I (7B) & CodeLlama-I (70B) & \starchat{} & GPT-3.5 & GPT-4 & \\

        \midrule

        \multirow{5}{*}{Java} & Reference & / & / & /& / & / & / & / & 86.15 & 1.42 & 2.62 & 2.58 & 3.08 & 2.8 & 3.19 \\

        & CodeLlama-I & 13.00 & 17.90 & 32.21 & 87.94 &59.61 & 86.88 & \textbf{50.69} & 46.62 & 2.32 & 3.72 & 2.80 & 3.28 & 3.64 & 3.93 \\

        & \starchat{} & \textbf{18.95} & \textbf{18.19} & \textbf{38.43} & \textbf{88.69} &\textbf{61.97} & \textbf{83.45} & 50.57 & \textbf{80.46} & 2.24 & 2.52 & 1.94 & 2.42 & 2.50 & 3.18 \\

        & GPT-3.5 & 12.49 & 16.74 & 31.87 &87.73 & 59.47 & 88.11 & 48.87 & 62.04 & 2.40 & 3.38 & 2.40 & 3.72 & 3.82 & 4.00 \\

        & GPT-4 & 9.46 & 17.02 & 28.36 & 86.72 & 58.83 & 89.27 & 46.50 & 36.12 & \textbf{2.44} & \textbf{3.88} & \textbf{2.60} & \textbf{4.10} & \textbf{4.50} & \textbf{4.17} \\

        \midrule
        
        \multirow{5}{*}{Python} & Reference & / & /& / & / & / & / & / & / & 1.48 & 3.13 & 2.74 & 2.84 & 2.98 & 3.56 \\

        & CodeLlama-I & 16.04 & \textbf{21.64} & 37.80 & \textbf{89.06} &\textbf{61.57} & \textbf{85.40} & \textbf{55.86} & / & 1.62 & 3.50 & 2.60 & 3.44 & 3.72 & 3.88 \\

        & \starchat{} & \textbf{18.35} & 17.62 & \textbf{37.96} & 88.92 &58.97 & 87.39 & 51.54 & / & 1.94 & 2.36 & 1.96 & 2.40 & 2.42 & 3.14 \\

        & GPT-3.5 & 11.95 & 19.14 & 30.20 & 87.63 &61.37 & 86.36 & 49.54 & / & \textbf{1.96} & \textbf{4.10} & \textbf{2.72} & \textbf{4.32} & \textbf{4.30} & \textbf{4.16} \\

        & GPT-4 & 14.07 & 20.87 & 35.38 & 88.11 &60.65 & 87.04 & 51.21 & / & 1.76 & 3.42 & 2.54 & 3.92 & 4.16 & 4.06 \\

        \midrule
        
        \multirow{5}{*}{C} & Reference & / & / & / & /& / & / & / & / & 1.56 & 2.25 & 2.80 & 2.24 & 2.62 & 3.05 \\

        & CodeLlama-I & 10.92 & \textbf{17.29} & 28.71 &86.38 & 51.55 &95.94 & 37.95 & / & 2.62 & 3.61 & \textbf{3.02} & 3.82 & 3.84 & 4.15 \\

        & \starchat{} & \textbf{15.58} & 15.57 & \textbf{32.84}  & \textbf{87.27} & \textbf{54.85} &\textbf{91.92} & \textbf{40.60} & / & 2.76 & 2.64 & 2.74 & 2.20 & 2.62 & 3.49 \\

        & GPT-3.5 & 12.06 & 16.00 & 29.81 &  86.65 &53.61 &93.71 & 39.75 & / & 3.04 & 3.38 & 2.86 & 3.48 & 3.66 & 4.06 \\

        & GPT-4 & 10.07 & 16.18 & 28.63 & 86.03 &53.00 & 94.77 & 37.30 & / & \textbf{3.18} & \textbf{3.72} & 2.86 & \textbf{4.00} & \textbf{4.36} & \textbf{4.25} \\
        
        \bottomrule
        
    \end{tabular}
    }
\end{table*}

\noindent\textbf{Correlation between Automated Evaluation and Human Evaluation.}
From Table~\ref{table:RQ1-all-metrics}, it can be observed that the average scores of reference summaries evaluated by the four LLM-based methods are mostly below 3 points. It means that similar to human evaluation, LLM-based evaluation methods also believe that the quality of the reference summaries is not very high. Besides, LLM-based evaluation methods are inclined to give higher scores to general-purpose LLMs GPT-3.5 and GPT-4, which is the same as human evaluation.

Based on the above observations, we can reasonably speculate that compared to methods based on summary-summary text/semantic similarity and summary-code semantic similarity, LLM-based evaluation methods may be more suitable for evaluating the quality of summaries generated by LLMs. 
Therefore, we follow~\cite{2022-Evaluation-Neural-Code-Summarization, 2021-Reassessing-Metrics-for-Code-Summarization} and calculate Spearman's correlation coefficient $\rho$ with the $p$-value between the results of each automated evaluation method and human evaluation, providing more convincing evidence for this speculation. 
The Spearman's correlation coefficient $\rho \in [-1,1]$ is suitable for judging the correlation between two sequences of discrete ordinal/continuous data, with a higher value representing a stronger correlation~\cite{1999-Practical-Nonparametric-Statistics}. $-1 \leq \rho < 0$, $\rho = 0$, and $0 < \rho \leq 1$ respectively indicate the presence of negative correlation, no correlation, and positive correlation~\cite{2007-Statistics-without-Maths-for-Psychology}. 
The $p$-value helps determine whether the observed correlation is statistically significant or simply due to random chance. 
By comparing the $p$-value to a predefined significance level (typically 0.05), we can decide whether to reject the null hypothesis and conclude that the correlation is statistically significant. 
Due to the page limit, we present the statistical results of $\rho$ and $p$-value in~\cite{2024-LLM4CodeSummarization}. 
The results demonstrate that among all automated evaluation methods, there is a significant positive correlation between the GPT-4-based evaluation method and human evaluation in scoring the quality of summaries generated by most LLMs, followed by the GPT-3.5-based evaluation method. 
For other automated evaluation methods, in most cases, their correlation with human evaluation is negative or weakly positive.
In a shot, existing metrics are inadequate for LLM-generated summaries. 
The main reason for this phenomenon lies in the low quality of reference summaries. The reference summaries are short and detail-lacking, while LLMs can generate detailed useful summaries not fully covered by the reference summaries. Therefore, the similarity with the reference summary cannot accurately reflect the quality of the LLM-generated summaries. 
In addition, BLEU penalizes long summaries, which also affects the correlation between BLEU and human evaluation. Although SIDE does not rely on reference summaries, it is trained on reference summaries and is therefore also affected. Re-training SIDE on LLM-generated summaries requires high-quality and extensive training data, while collecting large-scale LLM-generated summaries, especially from LLMs like GPT-4, is costly. Numerous factors influence LLM-generated summary quality, complicating the collection of high-quality data. 
Based on the above observations, we draw the conclusion that compared with other automated evaluation methods, the GPT-4-based method is more suitable for evaluating the quality of summaries generated by LLMs. In the subsequent RQs, we uniformly employ the GPT-4-based method to assess the quality of LLM-generated summaries. 
We set the temperature to default value 1 when using GPT-4 as the evaluator.

\summary{Among all automated evaluation methods, the GPT-4-based method overall has the strongest correlation with human evaluation. Therefore, it is recommended to adopt the GPT-4-based method to evaluate the quality of LLM-generated summaries.
}

\noindent\textbf{Why do LLMs give such a score?}

To investigate the scoring criteria of LLMs, we ask them to give explanations for their evaluation scores. The prompt we use is ``\textit{Here is a piece of code with corresponding comments and your previous rate for each comment on a scale from 1 to 5, where a higher score indicates better quality. Please explain why you scored this way.}''

\begin{figure}[!t]
    \centering
     \subfigure[A Java code snippet $c_1$]
    {
        \includegraphics[width=0.98\linewidth]{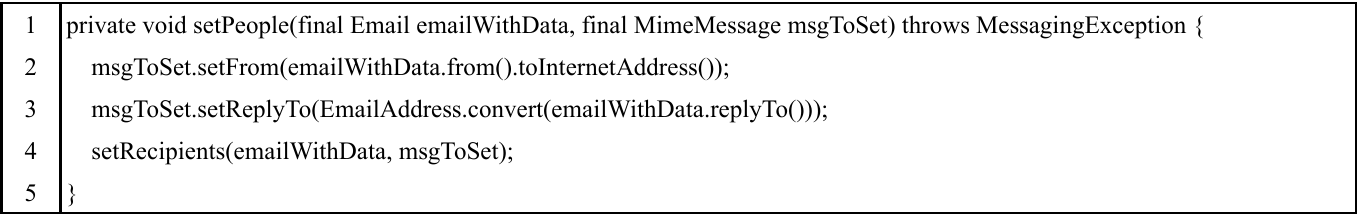}
        \label{fig:RQ1-explanation-code}
    }
    \subfigure[Reference summary and LLM-generated summaries for $c_1$]
    {
        \includegraphics[width=0.98\linewidth]{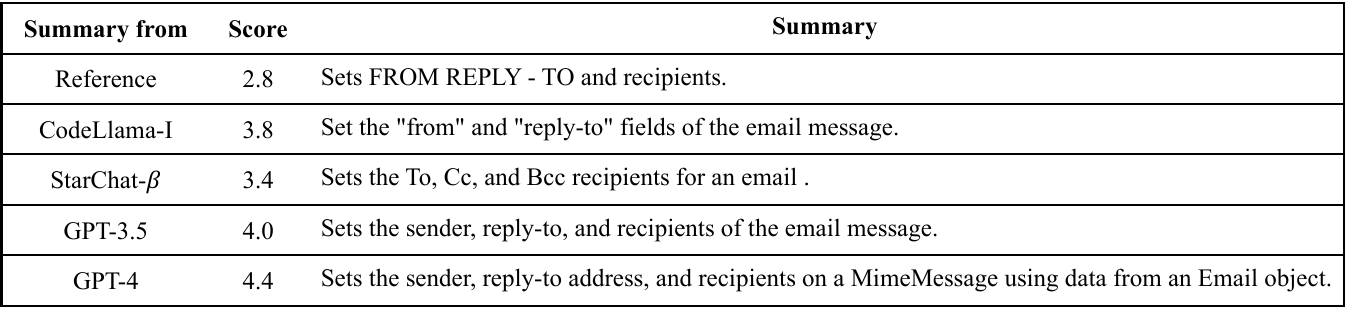}
        \label{fig:RQ1-explanation-summary}
    }
    \caption{An example of code, the reference summary, and LLM-generated summaries}
    \vspace{-4mm}
    \label{fig:RQ1-explanation-example}
\end{figure}

\begin{figure}[!t]
    \centering
    \subfigure[\codellama{} (70B)'s explanations]
    {
        \includegraphics[width=0.95\linewidth]{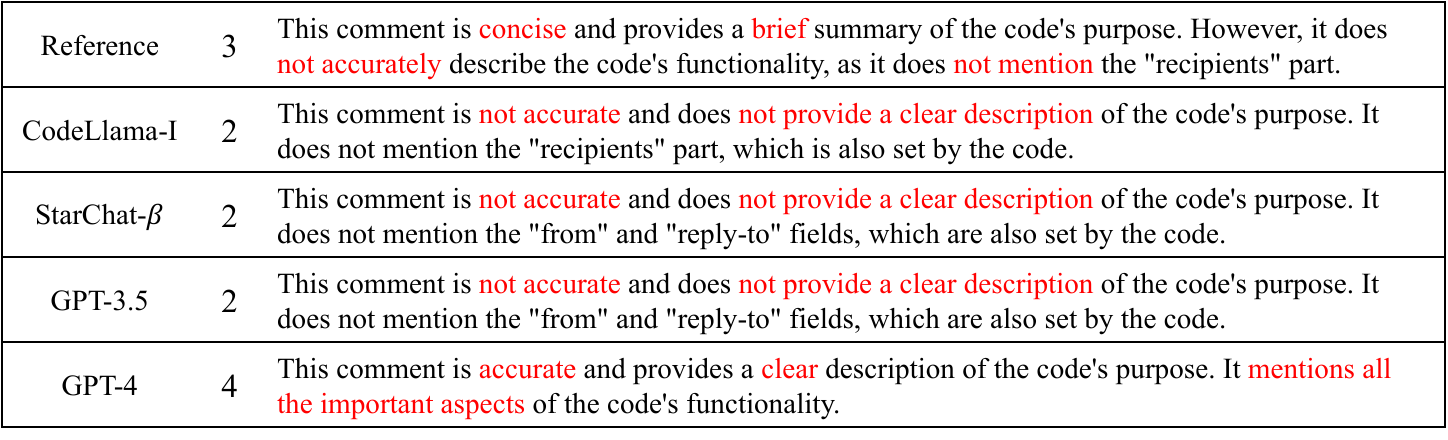}
        \label{fig:RQ1-explanation-codellama}
    }
    \subfigure[\starchat{}'s explanations]
    {
        \includegraphics[width=0.95\linewidth]{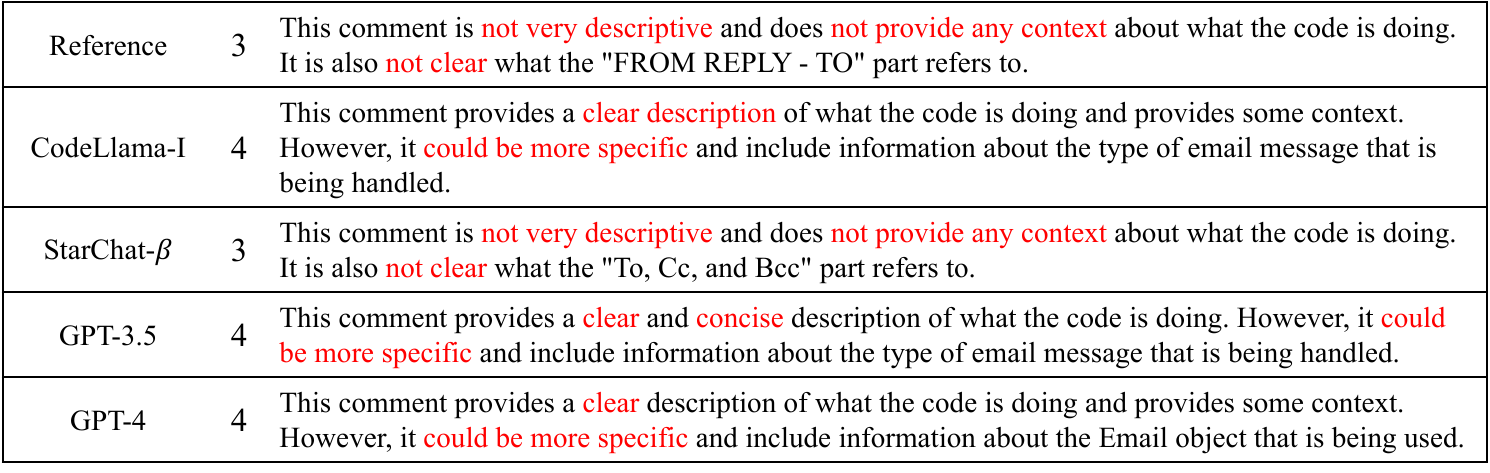}
        \label{fig:RQ1-explanation-starchat}
    }
    \subfigure[GPT-3.5's explanations]
    {
        \includegraphics[width=0.95\linewidth]{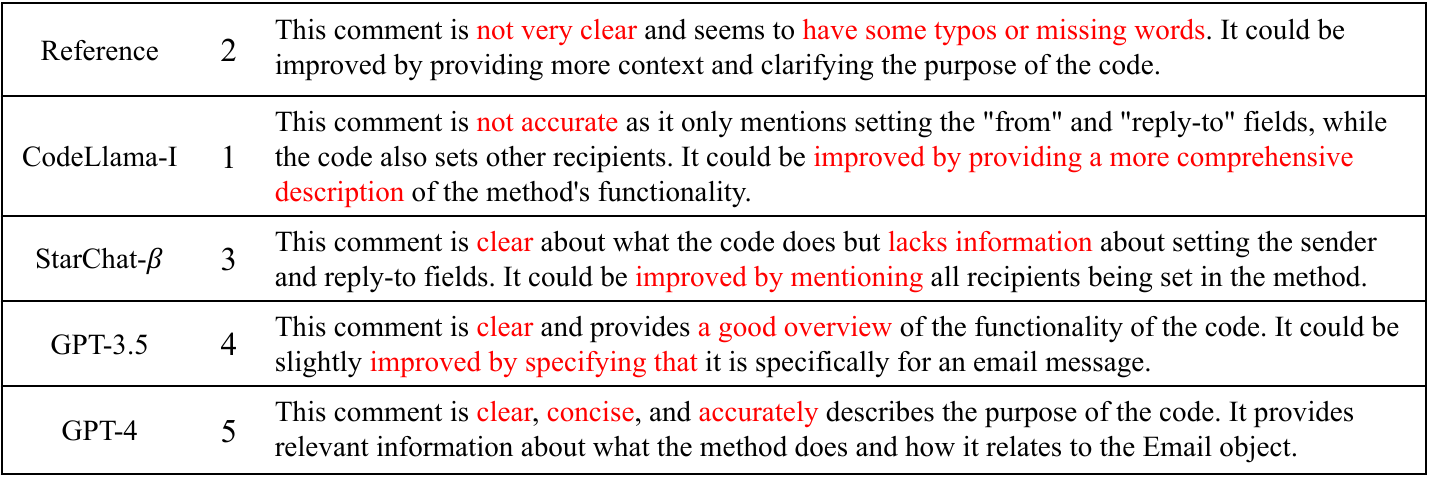}
        \label{fig:RQ1-explanation-gpt3.5}
    }
    \subfigure[GPT-4's explanations]
    {
        \includegraphics[width=0.95\linewidth]{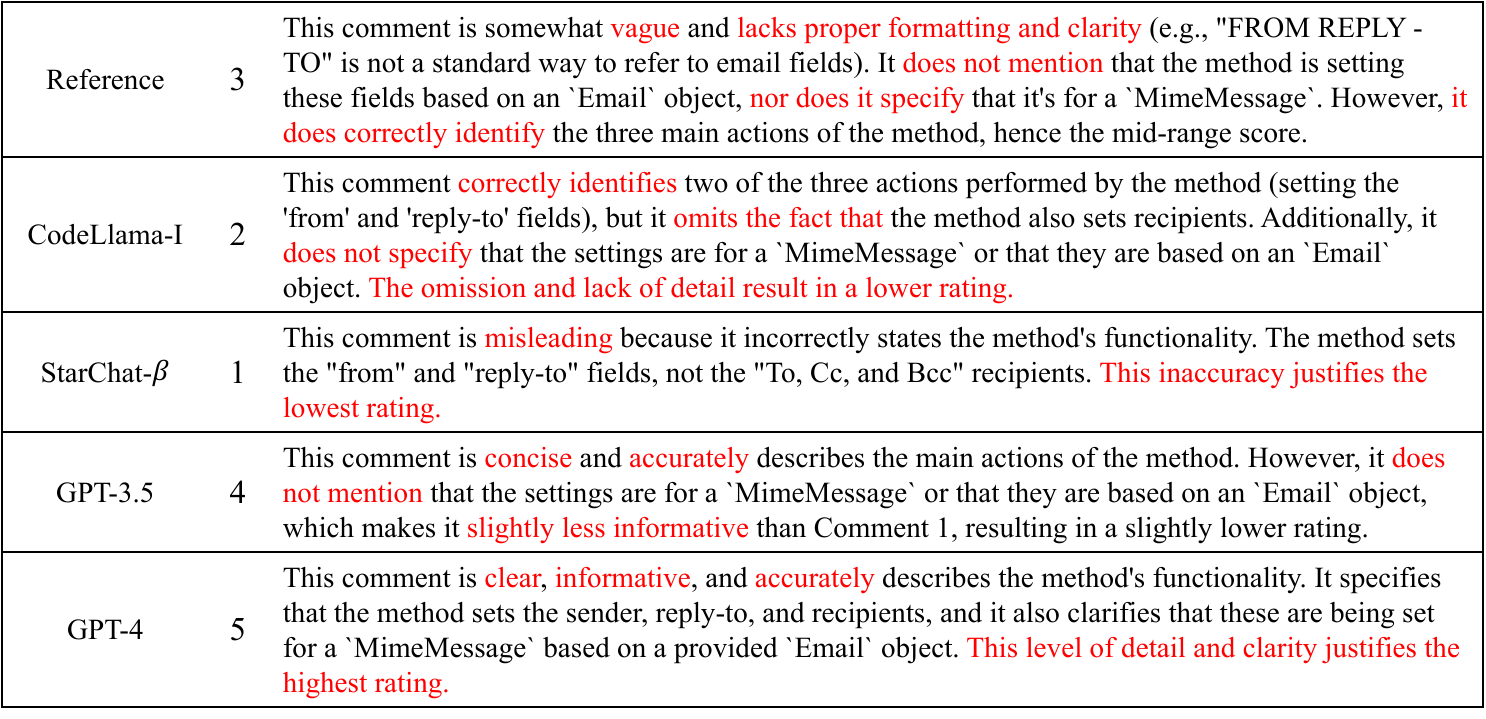}
        \label{fig:RQ1-explanation-gpt4}
    }
    \caption{LLMs' explanations of evaluation scores for the reference summary and LLM-generated summaries of $c_1$}
    \label{fig:RQ1-explanation}
\end{figure}

Figure~\ref{fig:RQ1-explanation} shows LLMs' explanations of evaluation scores for the reference summary and generated summaries of the example in Figure~\ref{fig:RQ1-explanation-example}. Due to the page limit, we only show the results of \codellama{} (70B), which performs better, instead of \codellama{} (7B). 
Observe that LLM's scoring criteria include accuracy, clarity, informativeness, and conciseness. 
Accuracy focuses on whether there is misleading information. 
Clarity focuses on whether the expression of a sentence is easy to understand. For example, the reference summary is evaluated by GPT-4 as not clear because ```FROM REPLY - TO' is not a standard way to refer to email fields''. 
Informativeness focuses on whether the summary contains enough details.
As for conciseness, since the summaries generated by GPT-3.5 and GPT-4 are also evaluated as concise, we can infer that LLMs' definition of concise is not equal to being as brief as the reference summary and the LLM-generated summaries are also considered concise. 
Therefore, the main factors affecting LLMs' scoring are accuracy, clarity, and informativeness. That is, whether there is misleading information, whether the expression is easy to understand, and whether the summary contains enough details. 
This also explains why LLMs do not give high scores to reference summaries. Reference summaries are generally short, do not cover every detail in the code snippet, and may contain terms in the code snippet which increases the difficulty of understanding. 
However, although the scoring criteria of LLMs are similar, the scoring results are still different. This is because LLMs' capabilities to understand code and summaries are different. As the SOTA LLM, GPT-4 achieves the best performance. It can accurately identify whether each summary contains all the details, while other LLMs may miss some of the details according to their explanation. In addition, GPT-4 does a better job in judging whether the summary is clear, as it points out that the phrase ``To, Cc, and Bcc'' is not easy to understand while GPT-3.5 and \codellama{} (70B) does not point out this problem.

\finding{The main factors affecting LLMs' scoring are accuracy, clarity, and informativeness. GPT-4 evaluation results have stronger correlation to human evaluation than other LLMs owing to its better understanding of code and summaries.}

\subsection{RQ2: How effective are different prompting techniques in adapting LLMs to the code summarization task?}
\label{subsec:answer_to_RQ2}

\noindent1) \textit{Experimental Setup.} The experimental dataset comprises 600 samples from Java, Python, and C datasets collectively.

\begin{table}[!t]
    \caption{Effectiveness of different prompting techniques}
    \label{table:RQ2-GPT-Eval}
    \centering
    \scriptsize
    \tabcolsep=8pt
    \begin{tabular}{ccccc}
        \toprule
        Model & Prompting Technique & Java & Python & C\\

        \midrule

        \multirow{5}{*}{\codellama{}}& zero-shot & 3.42 & 2.98 & 3.41 \\
        {} & few-shot & \textbf{3.78} & \textbf{3.75} & \textbf{3.91} \\
        {} & chain-of-thought &3.21 & 3.14 & 3.37 \\
        {} & critique & 2.15 & 2.02 & 2.13 \\
        {} & expert & 3.13 & 3.35 & 1.70 \\

        \midrule
        
        \multirow{5}{*}{\starchat{}}& zero-shot &2.71 & 2.85 & 2.86 \\
        {} & few-shot & 2.60 & 2.37 & 2.68 \\
        {} & chain-of-thought &\textbf{2.86} & 2.77 & \textbf{3.06}\\
        {} & critique &2.36 & 2.57 & 2.60 \\
        {} & expert &2.66 & \textbf{3.02} & 3.01 \\
        
        \midrule

        \multirow{5}{*}{GPT-3.5}& zero-shot & \textbf{3.90} & 3.96 & \textbf{3.93} \\
        {} & few-shot & 3.56 & \textbf{3.97} & 3.56  \\
        {} & chain-of-thought &3.36 & 3.47 & 3.36 \\
        {} & critique & 3.09 & 3.21 & 3.31 \\
        {} & expert & 2.72 & 3.43 & 3.49 \\

        \midrule
        
        \multirow{5}{*}{GPT-4}& zero-shot & 4.50 & 4.55 & 4.42 \\
        {} & few-shot & \textbf{4.66} & 4.16 & 4.18\\
        {} & chain-of-thought & 4.57 & \textbf{4.60} & 4.44\\
        {} & critique &4.41 & 4.44 & 4.34\\
        {} & expert &4.52 & 4.23 & \textbf{4.50}\\
        \bottomrule
    \end{tabular}
    \vspace{-2mm}
\end{table}

\noindent2) \textit{Experimental Results.}
Table~\ref{table:RQ2-GPT-Eval} presents the scores reported by the GPT-4 evaluation method for summaries generated by four LLMs using five prompting techniques. Observe that when the base model is \codellama{}, few-shot prompting consistently performs best on all three datasets. When the base model is \starchat{}, chain-of-thought prompting performs best on all the Java and C datasets, while expert prompting excels on the Python dataset. When selecting GPT-3.5 as the base model, the simplest zero-shot prompting surprisingly achieves the highest scores on the Java and C datasets, and is only slightly worse than few-shot prompting on the Python dataset. 
When using GPT-4 as the base model, chain-of-thought prompting overall performs best.

For the specific LLM and programming language, there is no guarantee that intuitively more advanced prompting techniques will surpass simple zero-shot prompting. 
For example, on the Java dataset, when selecting any of \starchat{}, GPT-3.5, and GPT-4 as the base model, few-shot prompting yields lower scores than zero-shot prompting. 
Contrary to the findings of previous studies~\cite{2022-Few-shot-Training-LLMs-for-Code-Summarization, 2024-LLM-Few-Shot-Summarizers-Multi-Intent-Comment-Generation}, the GPT-4-based evaluation method does not consider that few-shot prompting will improve the quality of generated summaries. This discrepancy may arise because previous studies evaluated the quality of LLM-generated summaries using BLEU, METEOR, and ROUGE-L, which primarily assess text/semantic similarity with reference summaries. However, as we mentioned in Section~\ref{subsec:answer_to_RQ1}, reference summaries contain low-quality noisy data that undermines their reliability. Therefore, achieving greater similarity with reference summaries does not necessarily imply that the human/GPT-4-based evaluation method will perceive the summary to be of higher quality. 

\summary{The more advanced prompting techniques expected to perform better may not necessarily outperform simple zero-shot prompting. In practice, selecting the appropriate prompting technique requires considering the base LLM and the programming language.}

\subsection{RQ3: How do different model settings affect LLMs' code summarization performance?}
\label{subsec:answer_to_RQ3}

\noindent1) \textit{Experimental Setup.} There are three key model settings/parameters, including top\_k, top\_p, and temperature, that allow the user to control the randomness of text (code summary in our scenario) generated by LLMs. Considering that GPT-3.5 and GPT-4 do not support the top\_k setting, we only conduct experiments with the top\_p and temperature. 

\noindent\textbf{Top\_p:} In each round of token generation, LLMs sort tokens by probability from high to low and keep tokens whose probability adds up to (no more than) top\_p. For example, $top\_p=0.1$ means only the tokens comprising the top 10\% probability mass are considered. 
The larger the top\_p is, the more tokens are sampled. Thus tokens with low probabilities have a greater chance of being selected, so the summary generated by LLMs is more random.

\noindent\textbf{Temperature:} 
Temperature adjusts the probability of tokens after top\_p sampling. 
The higher the temperature, the less the difference between the adjusted token probabilities. Therefore, the token with a low probability has a greater chance of being selected, so the generated summary is more random. If the temperature is set to 0, the generated summary is the same every time.

Top\_p and temperature are alternatives and one should only modify one of the two parameters at a time~\cite{2024-Create-Chat-Completion}. Therefore, the questions we want to answer are: (1) Does top\_p/temperature impact the quality of LLM-generated summaries? (2) As alternative parameters that both control the randomness of LLMs, do top\_p and temperature have a difference in the degree of influence on the quality of LLM-generated summaries?

Drawing from a review of related work (see Section~\ref{sec:background_and_related_work}), we find that existing LLM-based code summarization studies pay more attention to few-shot prompting. Since no prompting technique outperforms others on all LLMs, we uniformly employ few-shot prompting in RQ3, RQ4, and RQ5 to facilitate comparing our findings with prior studies.

\begin{table}[t]
    \caption{Influence of different model settings. 
    We bold the scores of the best setting combinations on each dataset.}
    \label{table:RQ3-GPT-Eval}
    \centering
    \scriptsize
    \tabcolsep=8pt
    \begin{tabular}{cccccc}
        \toprule
        Model & Top\_p & Temperature & Java & Python & C \\
        
        \midrule

        \multirow{9}{*}{\codellama{}} & \multirow{3}{*}{0.5} & 0.1 &3.81 &3.83 &4.10 \\
        {} & {} & 0.5 & 3.72 & 3.85 &4.08 \\
        {} & {} & 1.0 &\textbf{3.91} &3.81 &\textbf{4.11} \\
        
        \cmidrule{2-6}
        
        {} & \multirow{3}{*}{0.75} & 0.1 & 3.76  &\textbf{3.87} & 4.02 \\
        
        {} & {} & 0.5 & \textbf{3.91} & 3.73 & 4.01 \\
        
        {} & {} & 1.0 & 3.80 & 3.79 &3.88 \\
        
        \cmidrule{2-6}
        
        {} & \multirow{3}{*}{1.0} & 0.1 &  3.78 & 3.75 & 3.91\\
        
        {} & {} & 0.5 & \textbf{3.91} & 3.75 & 3.99 \\
        
        {} & {} & 1.0 & 3.73 & 3.59 &3.60\\
                            
        \midrule
        
        \multirow{9}{*}{\starchat{}} & \multirow{3}{*}{0.5} & 0.1 & 2.49 & 2.42 & 2.72 \\
        
        {} & {} & 0.5 & 2.47 & 2.36 &2.70 \\
        
        {} & {} & 1.0 & 2.49 &2.29 &2.75 \\
        
        \cmidrule{2-6}
        
        {} & \multirow{3}{*}{0.75} & 0.1 & 2.50 & 2.35 & 2.66 \\
        
        {} & {} & 0.5 &  2.45 &\textbf{2.47} & \textbf{2.80} \\
        
        {} & {} & 1.0 &2.48 &2.37 &2.71 \\
        
        \cmidrule{2-6}
        {} & \multirow{3}{*}{1.0} & 0.1 &\textbf{ 2.60} & 2.37  & 2.68\\
        {} & {} & 0.5 & 2.53 &2.45 &2.77 \\
        {} & {} & 1.0 & 2.54 &2.38 &2.69\\
                            
        \midrule
        
        \multirow{9}{*}{GPT-3.5} & \multirow{3}{*}{0.5} & 0.1 &3.41 &3.60 &3.40 \\
        {} & {} & 0.5 & 3.45 &3.73 &3.38 \\
        {} & {} & 1.0 & 3.52 &3.68 &3.42 \\
        \cmidrule{2-6}
        {} & \multirow{3}{*}{0.75} & 0.1 &3.54 &3.66&3.35 \\
        {} & {} & 0.5 & 3.55 &3.65 &3.24 \\
        {} & {} & 1.0 & 3.46 &3.64 &3.41 \\
        \cmidrule{2-6}
        {} & \multirow{3}{*}{1.0} & 0.1 & \textbf{3.56} & \textbf{3.97} & \textbf{3.56}\\
        {} & {} & 0.5 & 3.55 &3.71 &3.42 \\
        {} & {} & 1.0 & 3.41 &3.72 &3.52\\
        
        \midrule
        
        \multirow{9}{*}{GPT-4} & \multirow{3}{*}{0.5} & 0.1 &4.44 &4.25 &4.33 \\
        {} & {} & 0.5 & 4.47 &4.30 &4.31 \\
        {} & {} & 1.0 & 4.45 &4.31 &4.29 \\
        \cmidrule{2-6}
        {} & \multirow{3}{*}{0.75} & 0.1 &4.48 &4.27&4.31 \\
        {} & {} & 0.5 & 4.46 &\textbf{4.34} &4.26 \\
        {} & {} & 1.0 & 4.47 &4.33 &\textbf{4.36} \\
        \cmidrule{2-6}
        {} & \multirow{3}{*}{1.0} & 0.1 & \textbf{4.66} & 4.16 & 4.18\\
        {} & {} & 0.5 & 4.43 &4.27 &4.33 \\
        {} & {} & 1.0 &4.40 &4.18 &4.33 \\
        
        \bottomrule
        
    \end{tabular}
    \vspace{-5mm}
\end{table}

\noindent2) \textit{Experimental Results.} 
TABLE~\ref{table:RQ3-GPT-Eval} shows the scores evaluated by the GPT-4 evaluation method for the summaries generated by LLMs under different top\_p and temperature settings. 
It is observed that the impact of top\_p and temperature on the quality of LLM-generated summaries is specific to the base LLM and programming language.  
For example, when top\_p=0.5, as temperature increases, the quality of GPT-4-generated summaries for Python code snippets increases, while those for C code snippets decrease. Another example is that when top\_p=0.5, as the temperature rises, the quality of GPT-4-generated Java comments first increases and then decreases, whereas \codellama{} is exactly the opposite, first decreases and then increases. 
Regarding the difference in influence between top\_p and temperature, it is observed that in most cases the influence of the two parameters is similar. For example, for C code snippets, when one parameter (top\_p or temperature) is fixed, as the other parameter (temperature or top\_p) grows, the quality of GPT-3.5-generated summaries first decreases and then increases.

\summary{The impact of top\_p and temperature on the quality of generated summaries is specific to the base LLM and programming language. As alternative parameters, top\_p and temperature have similar influence on the quality of LLM-generated summaries. The impact of top\_p and temperature on GPT-4 is small.}

\subsection{RQ4: How do LLMs perform in summarizing code snippets written in different types of programming languages?}
\label{subsec:answer_to_RQ4}

\noindent1) \textit{Experimental Setup.} 
We conduct experiments on all 10 programming language datasets. 
As in RQ3, we uniformly employ few-shot prompting to adapt LLMs. 

\begin{table}[!t]
    \caption{Effectiveness of LLMs in summarizing code snippets written in different types of programming languages. 
    CodeLlama-I: \codellama{}.
    }
    \label{table:RQ4-GPT-Eval}
    \centering
    \scriptsize
    \tabcolsep=1.5pt
    \begin{tabular}{ccccccccccc}
        \toprule
        
        \multirow{2}{*}{Model} & OOP & \multicolumn{2}{c}{PP} & \multicolumn{4}{c}{SP} & \multicolumn{2}{c}{FP} & LP \\

        \cmidrule(lr){2-2}\cmidrule(lr){3-4}\cmidrule(lr){5-8}\cmidrule(lr){9-10}\cmidrule(lr){11-11}
        
        & Java & C & Go & Python & Ruby & PHP & JavaScript & Erlang & Haskell & Prolog \\
        
        \midrule
        
        CodeLlama-I & 3.78 & 3.91 & 3.86 & 3.75 & 3.98 &3.88 &4.03 &3.51 &3.58 &3.23 \\
       
        \starchat{} & 2.60 & 2.68 & 2.97 & 2.37 & 2.79 & 2.73 & 2.67 & 2.68 & 2.88 & 2.34 \\
        
        GPT-3.5 & 3.56 & 3.56 & 4.14 & 3.97 & 3.64 & 3.99 & 3.53 & 3.57 & 3.44 & 3.42 \\
        
        GPT-4 & \textbf{4.66} & \textbf{4.18} & \textbf{4.36} & \textbf{4.16} & \textbf{4.37} & \textbf{4.31} & \textbf{4.29} & \textbf{4.23} & \textbf{4.22} & \textbf{4.05} \\
        
        \bottomrule
    \end{tabular}
    \vspace{-3mm}
\end{table}

\noindent2) \textit{Experimental Results.}
Table~\ref{table:RQ4-GPT-Eval} shows the performance evaluated by the GPT-4 evaluation method for the four LLMs on five types of programming languages. It is observed that 
for OOP (i.e., Java), GPT-4 performs best, followed by \codellama{}, GPT-3.5, and \starchat{}. 
For PP, GPT-4 performs best on both C and Go, while \starchat{} performs worst on both. The smallest LLM \codellama{} outperforms GPT-3.5 on C (3.91 vs. 3.56), but vice versa on Go (3.86 vs. 4.14). Additionally, except for \codellama{}, which performs slightly worse on Go than on C (3.86 vs. 3.91), the other three LLMs perform better on Go than on C. 
For SP, GPT-4 consistently performs best on all four languages. Surprisingly, \codellama{} outperforms GPT-3.5 on both Ruby and JavaScript. All four LLMs perform better on PHP than on Python. 
For FP, the performance of two specialized code LLMs (i.e., \codellama{} and \starchat{}) is better on Haskell than on Erlang, while the opposite is true for the two general-purpose LLMs (i.e., GPT-3.5 and GPT-4). 
For LP, GPT-4 still performs best, followed by GPT-3.5, \codellama{}, and \starchat{}. 
Across all five types of languages, the four LLMs consistently perform the worst on LP, which indicates that summarizing logic programming language code is the most challenging. One possible reason is that fewer Prolog datasets are available for training these LLMs compared to other programming languages. The scale of the Prolog dataset we collected can support this reason.

\summary{GPT-4 surpasses the other three LLMs on all five types of programming languages. For PP, LLMs overall perform better on Go than on C. For SP, all four LLMs perform better on PHP than on Python. For FP, specialized code LLMs (e.g., \starchat{}) perform better on Haskell than on Erlang, whereas the reverse is true for general-purpose LLMs (e.g., GPT-4). All four LLMs perform worse in summarizing LP code snippets.}

\subsection{RQ5: How do LLMs perform on different categories of summaries?}
\label{subsec:answer_to_RQ5}

\noindent1) \textit{Experimental Setup.}
Following~\cite{2020-CPC, 2021-Why-My-Code-Summarization-Not-Work, 2023-DOME}, we classify code summaries into the following six categories.

\noindent\textbf{What}: 
describes the functionality of the code snippet. It helps developers to understand the main functionality of the code without diving into implementation details. An example is ``Pushes an item onto the top of this stack''.

\noindent\textbf{Why}: 
explains the reason why the code snippet is written or the design rationale of the code snippet. It is useful when methods' objective is masked by complex implementation. 
An application scenario of \texttt{Why} summaries is to explain the design rationale of overloaded functions.

\noindent\textbf{How-it-is-done}:  
describes the implementation details of the code snippet. Such information is critical for developers to understand the subject, especially when the code complexity is high. For instance, ``Shifts any subsequent elements to the left." is a \texttt{How-it-is-done} comment. 

\noindent\textbf{Property}: 
asserts properties of the code snippet, e.g., function's pre-conditions/post-conditions.  
``This method is not a constant-time operation." is a \texttt{Property} summary. 

\noindent\textbf{How-to-use}: 
describes the expected set-up of using the code snippet, such as platforms and compatible versions. For example, ``This method can be called only once per call to next()." is a \texttt{How-to-use} summary. 

\noindent\textbf{Others}: Comments that do not fall into the above five categories are classified as \texttt{Others} summaries, such as ``The implementation is awesome.". 
Following Mu et al.~\cite{2023-DOME}, we consider the $\langle code, summary\rangle$ pairs with \texttt{Others} comments as noisy data, and remove them if identified.

\begin{table}[t]
    \caption{Statistics of six sub-datasets divided from the CSN-Java test dataset according to comment intention}
    \label{table:RQ5-subsets}
    \centering  
    \scriptsize
    \tabcolsep=16pt
    \begin{tabular}{ccc}
        \toprule
        Summary Category & Number of Samples & Sample Ratio \\
        
        \midrule
        
        What &  6,132 & 0.56 \\
        
        Why & 1,190 & 0.11 \\
        
        How-it-is-done & 2,242 & 0.20 \\
        
        Property & 1,174 & 0.11 \\
        
        How-to-use & 180 & 0.02 \\
        
        Others & 37 & $<0.01$ \\ 
        \bottomrule
    \end{tabular}
\end{table}

We employ the comment classifier COIN provided by Mu et al.~\cite{2023-DOME} to classify the CSN-Java dataset according to the comment intention type. The test dataset is divided into six sub-datasets, as shown in Table~\ref{table:RQ5-subsets}. 
To facilitate comparison between different categories, we randomly select 180 samples from each sub-dataset. 
As in RQ4, we uniformly employ few-shot prompting to adapt LLMs. For each sub-dataset with different intention types, the few-shot example is of the same intention type from the training dataset.

\begin{table}[t]
    \caption{Effectiveness of LLMs in generating different categories of summaries}
    \label{table:RQ5-GPT-Eval}
    \centering
    \scriptsize
    \tabcolsep=5pt
    \begin{tabular}{cccccc}
        \toprule
        Model & What & Why & How-it-is-done & Property & How-to-use\\

        \midrule

        \codellama{} & 4.15  & \textbf{4.29}  & 3.85  & \textbf{4.19}  & 3.96 \\
        
        \starchat{} & 2.68  & 2.78  & 2.77  & 2.94  & 2.52 \\

        GPT-3.5 & 3.61  & 3.54  & 3.97  & 3.54  & 4.17 \\

        GPT-4 &\textbf{4.40}  & 4.28  & \textbf{4.31}  & 4.06  & \textbf{4.22} \\
        
        \bottomrule
        
    \end{tabular}
    \vspace{-2mm}
\end{table}

\noindent2) \textit{Experimental Results.}
Table~\ref{table:RQ5-GPT-Eval} presents the results evaluated by the GPT-4 evaluation method for the four LLMs in generating five categories of summaries. Observe that \codellama{} performs worse in generating \texttt{How-it-is-done} summaries than generating the other four categories of summaries. 
\starchat{} gets the lowest score of 2.52 in generating \texttt{How-to-use} summaries. 
Both GPT-3.5 and GPT-4 are not as good at generating \texttt{Property} summaries compared to generating other categories of summaries. 
Surprisingly, the smallest LLM \codellama{} slightly outperforms the advanced GPT-4 in generating \texttt{Why} (4.29 vs. 4.28) and \texttt{Property} (4.19 vs. 4.06) summaries. 
Additionally, compared with GPT-3.5, \codellama{} achieves higher scores in generating \texttt{What}, \texttt{Why}, and \texttt{Property} summaries. 
Certainly, it is undeniable that the reason for this phenomenon is that the optimal prompting technique for GPT-3.5 and GPT-4 is not few-shot prompting. This phenomenon is also exciting because it implies that most ordinary developers or teams who lack sufficient resources (e.g., GPUs) have the opportunity to utilize open-source and small-scale LLMs to achieve code summarization capabilities close to (or even surpass) those of commercial gigantic LLMs.

\summary{The four LLMs excel in generating different categories of summaries. The smallest \codellama{} slightly outperforms the advanced GPT-4 in generating \texttt{Why} and \texttt{Property} summaries. \starchat{} is not proficient at generating \texttt{How-to-use} summaries. GPT-3.5 and GPT-4 perform worse in generating \texttt{Property} summaries than other categories of summaries.}

\section{Threats to Validity}
\label{sec:threats_to_validity}
Our empirical study may contain several threats to validity that we have attempted to relieve.

\textbf{Threats to External Validity.} The threats to external validity lie in the generalizability of our findings. One threat to the validity of our study is that LLMs usually generate varied responses for identical input across multiple requests due to their inherent randomness, while conclusions drawn from random results may be misleading. To mitigate this threat, considering that \starchat{} and \codellama{} do not support setting the temperature to 0, we uniformly set it to 0.1 to reduce randomness except for RQ3. In RQ2-RQ5, to make the evaluation scores more deterministic, we set the temperature to 0 when using GPT-4 as the evaluator. 
Additionally, for other RQs, we conduct experiments on multiple programming languages to support our findings.

\textbf{Threats to Internal Validity.} 
A major threat to internal validity is the potential mistakes in the implementation of metrics and models. To mitigate this threat, we use the publicly available code from previous studies~\cite{2020-Rencos, 2023-SIDE} for BLEU, METEOR, ROUGE-L, and SIDE. 
For COIN, BERTScore, SentenceBert, Universal Sentence Encoder, \starchat{}~\cite{2023-Replication-Package-for-StarChat-beta} and \codellama{}~\cite{2023-Replication-Package-for-CodeLlama}, and GPT-3.5/GPT-4~\cite{2024-Replication-Package-for-GPT-API}, we use the script provided along with the model to run.

Another threat lies in the processing of LLM’s responses. Usually, the output of LLMs is a paragraph, not a sentence of code summary (code comment) that we want. The real code summary may be the first sentence in the LLMs' response, or it may be returned in the comment before the code such as ``/** $\langle$code summary$\rangle$ */", etc. Therefore, we designed a series of heuristic rules to extract the code summary. We have made our script for extracting code summaries from LLMs' responses public for the community to review.

\section{Conclusion}
\label{sec:conclusion}
In this paper, we provide a comprehensive study covering multiple aspects of code summarization in the era of LLMs. Our interesting and significant findings include, but are not limited to, the following aspects. 1) Compared with existing automated evaluation methods, the GPT-4-based evaluation method is more fitting for assessing the quality of LLM-generated summaries. 2) The advanced prompting techniques anticipated to yield superior performance may not invariably surpass the efficacy of straightforward zero-shot prompting. 3) The two alternative model settings have a similar impact on the quality of LLM-generated summaries, and this impact varies by the base LLM and programming language. 4) LLMs exhibit inferior performance in summarizing LP code snippets. 5) CodeLlama-Instruct with 7B parameters demonstrates superior performance over the advanced GPT-4 in generating \texttt{Why} and \texttt{Property} summaries. 
Our comprehensive research findings will aid subsequent researchers in swiftly grasping the various facets of LLM-based code summarization, thereby promoting the development of this field.

\section*{Acknowledgment}
The authors would like to thank the anonymous reviewers for their insightful comments. 
This work is supported by the National Research Foundation, Singapore, and DSO National Laboratories under the AI Singapore Programme (AISG Award No: AISG2-GC-2023-008), the National Research Foundation, Singapore, and the Cyber Security Agency under its National Cybersecurity R\&D Programme (NCRP25-P04-TAICeN), and the National Natural Science Foundation of China (61932012, 62372228). Any opinions, findings and conclusions or recommendations expressed in this material are those of the author(s) and do not reflect the views of the National Research Foundation, Singapore and Cyber Security Agency of Singapore. Chunrong Fang is the corresponding author.

\bibliographystyle{IEEEtran}
\bibliography{reference}

\balance


\end{document}